# Coexistence of spontaneous polarization and superconductivity in hole-doped oxyhydrides ATiO$_2$H (A=K, Rb, Cs): first-principles study

Minjae Ghim[1,*], Nobuya Sato[2], Ryosuke Akashi[2], Seung-Hoon Jhi[1], and Shinji Tsuneyuki[2,3]

[1]Department of Physics, Pohang University of Science and Technology, Pohang, 37673, Republic of Korea

[2]Department of Physics, The University of Tokyo, 7-3-1 Hongo, Bunkyo-ku, Tokyo 113-0033, Japan

[3]The Institute for Solid State Physics, The University of Tokyo, 5-1-5 Kashiwanoha, Kashiwa-shi, Chiba 277-8581, Japan

* mjkim1790@gmail.com

ABSTRACT: The polar metal is a material that hosts both polar distortion and metallicity. Such a material is expected to show exotic magneto-electric phenomena if superconducts. Here, we theoretically explore ferroelectric and superconducting properties in a series of perovskite-type oxyhydrides ATiO$_2$H (A=K, Rb, Cs) under hole-doping conditions using the first-principles calculations based on the density functional theory. Our simulation shows that these compounds host spontaneous polarization and superconductivity at optimal doping concentration. The unusual coexistence of superconductivity and large polarization (~100 μC/cm$^2$) originates from weak coupling of the polar distortion and superconducting states, the reason of which is separation of the displaced atoms and spacially confined metallic carriers. Besides, the superconductivity is enhanced by the unique electronic properties near the valence band maximum: quartic band dispersion with a sizable contribution of hydrogen 1$s$ states. Our study thus feature the oxyhydrides as possible model polar superconducting systems, which may be utilizable for future magneto-electric devices.

# I. INTRODUCTION

A polar metal exhibits both polar distortion and metallicity. Such a metal is uncommon in reality because itinerant carriers screen the long-range electrostatic interaction. Once combined with superconductivity, the material is expected to show exotic physical properties such as the magneto-electric effect [1,2], different upper critical fields for opposite direction [3] and unconventional Cooper paring, as a non-centrosymmetric superconductor [4,5]. The concept of the polar metal was firstly suggested by Anderson and Blout [6]. Its strict criteria involve not only the coexistence of the metallicity and polar distortion but also the continuous onset of the distortion [7]. The first discovered polar metal satisfying those criteria is $LiOsO_3$ [8]. Since then, several materials are also found to be candidate polar metals including native polar semimetal $WTe_2$ [9,10] and $MoTe_2$ [11,12], and doped polar metals $BaTiO_3$ [13,14] and $SrTiO_3$ [15]. Especially, $MoTe_2$ [16,17] and doped $SrTiO_3$ [15] are considered as ferroelectric metals with superconductivity.

Regardless if the polar distortion is continuous or not, the coexistence of polar distortion and metallicity itself can be the source of the abovementioned exotic phenomena and has been readdressed in several studies [18–20]. The metallicity is usually detrimental to the polar distortion, but it is so when long-range electrostatic interaction is mostly involved in the polar distortion. Metal may have polar distortion if it is caused by local bonding [18], geometric origin [20], or when the polar phonon mode and itinerant-carriers are weakly coupled as in $LiOsO_3$ [19,21].

In this context, we propose that a series of perovskite oxyhydrides $ATiO_2H$ (A=K, Rb, Cs) polar distortion and metallicity can coexist under hole-doping conditions, and that superconducting phase with the polar distortion emerges at optimal hole-doping concentration. Recently several kinds of stochiometric oxyhydrides such as BaScO$_2$H and Ba$_2$YHO$_3$ have been synthesized [22–26]. On top of this, ATiO$_2$H have been predicted as a system hosting several interesting electronic/ionic properties with a possible chemical synthesis route [27,28]. These compounds are perovskite-type oxyhydrides with large spontaneous polarization of about 100 μC/cm$^2$. These compounds have several electronic properties advantageous for the persisting polar distortion and the superconductivity under hole-doping conditions. The first advantage is that the polar distortion is not caused by the long-range interaction. The polar distortion is mainly due to the displacement of the apical oxygen (O$_{ap}$) along the c-axis [Fig 1(a)]. The $zz$ component of the Born effective charge tensor of O$_{ap}$ is only about $-2$ in ATiO$_2$H [28], much smaller in its size than $+5.8$ and $-4.7$ of Ti and O$_{ap}$ $zz$ component, respectively, in the tetragonal phase of BaTiO$_3$ [29]. The small Born effective charge indicates the minor contribution of the long-range Coulomb interaction to the polar distortion in ATiO$_2$H. Meanwhile, in BaTiO$_3$, the Coulomb interaction gives a major contribution to the tetragonal phase, which is responsible for the fragility of the phase with electron doping [14]. The second advantage is that the polar distortion is weakly coupled to the itinerant carriers. In Fig 2(b), O$_{ap}$-$2p$ states are far below from the valence band maximum (VBM). Therefore, the polar distortion may be resistant to the hole-doping as in LiOsO$_3$. Third, the oxyhydrides show two-dimensional (2D) electronic band structures near the VBM, the dispersionless band along the Y-T direction. The 2D electronic structure shows the step-like density of states (DOS) near the VBM, and even small hole-doping can provide sizable DOS at Fermi-level. Such a 2D property leads to high superconducting transition temperature ($T_c$) of graphane C$_2$H$_2$ (prediction) and MgB$_2$ even at small doping or charge transfer [30]. Moreover, the hydrogen (H)-1$s$ orbital gives sizable contribution to the states near the VBM. The hydrogen vibrational modes can contribute to the superconductivity even in ambient pressure and may enhance the superconductivity as in the pressurized hydride superconductors [31,32].

In this paper, we studied ATiO$_2$H (A=K, Rb, Cs) as candidate systems that concurrently host polarization and superconductivity using the density functional theory (DFT). In particular, we pursued the optimal hole-doping conditions and structures for superconductivity and polarization.

This paper is organized as follows: In Sec. II, we introduce a scheme to estimate spontaneous polarization in 2D-type metals, and schemes to improve convergence problems in calculating $T_c$ of 2D-type superconductors. In Sec. III, under hole-doping conditions, the optimized structures, spontaneous polarization, phonon dispersion, and superconductivity of the ATiO$_2$H are presented and discussed. Sec. IV summarizes our conclusion.

## II. METHOD

A series of hydroxides ATiO$_2$H (A=K, Rb, Cs) have 2D-like electronic structures near the Fermi-level under hole-doping conditions. The 2D-like feature, on one hand, gives us concrete definition of the spontaneous polarization in metals and, on the other hand, is difficult to treat for accurate estimation of the superconducting transition temperature ($T_c$). In this section, we introduce methods to calculate the polarization of 2D-type metals and also $T_c$.

A. Calculation of spontaneous polarization of 2D-type metals.

In the Berry phase scheme [33,34], the electronic contribution to the change in polarization is expressed as

$$\Delta P_\text{e} = P_\text{e}(\theta_2) - P_\text{e}(\theta_1) \qquad (1)$$

with

$$P_\text{e}(\theta) = -\frac{2e}{(2\pi)^3} \int_{A_\mathbf{k}} d\mathbf{k}_\perp \phi^\theta(\mathbf{k}_\perp), \qquad (2)$$

where $\theta$ is an adiabatic parameter corresponding to a crystal structure of an adiabatic transition path, and $A_\mathbf{k}$ is the area perpendicular to the polarization direction inside Brillouin zone. $\phi^\theta(\mathbf{k}_\perp)$ is the Berry phase at $\mathbf{k}_\perp \in A_\mathbf{k}$,

$$\phi^\theta(\mathbf{k}_\perp) = \lim_{N\to\infty} \text{Im}\left\{\ln \prod_{j=0}^{N-1} \det\left\langle u^\theta_{m\mathbf{k}_j} \middle| u^\theta_{n\mathbf{k}_{j+1}}\right\rangle\right\}, \qquad (3)$$

where $u^\theta_{n\mathbf{k}_j}$ is a cell-periodic part of a wavefunction. The band indices $m$ and $n$ run over occupied bands, and $\mathbf{k}_j = \mathbf{k}_\perp + jG_\parallel/N$, where $G_\parallel$ is the reciprocal vector along the polarization direction. For insulators, Eq. (3) is cyclic and gauge-invariant, and the polarization is well-defined. Meanwhile, in metals, the Berry phase is gauge-dependent and the polarization is ill-defined because the occupation can change along the polarization direction.

However, in 2D-type metals, the occupation changes slightly along the polarization direction, and the polarization can be well-defined. Utilizing this advantage, a modified Berry phase method is introduced for 2D-type ferroelectric metals [35]. In this scheme, the polarization is separated into $P_e^{\text{bare}}$, a bare polarization assuming that each band is fully occupied or unoccupied, and $P_e^{\text{ep}}$ ($P_e^{\text{hp}}$), the contribution of electron (hole) pocket to the polarization.

$$P_e(\theta) = P_e^{\text{bare}}(\theta) + P_e^{\text{ep}}(\theta) - P_e^{\text{hp}}(\theta), \qquad (4)$$

where $P_e^{\text{bare}}$ is obtained by the original Berry phase scheme, and $P_e^{\text{ep}}$ ($P_e^{\text{hp}}$) is obtained by confining the integration area $A_\mathbf{k}$ of Eq. (2) to the area of electron (hole) pocket $A_\mathbf{k}^{\text{ep(hp)}}$:

$$P_e^{\text{ep(hp)}}(\theta) = -\frac{2e}{(2\pi)^3} \int_{A_\mathbf{k}^{\text{ep(hp)}}} d\mathbf{k}_\perp \phi^\theta(\mathbf{k}_\perp). \qquad (5)$$

This scheme was successful in reproducing the measured spontaneous polarization of bulk WTe$_2$ [9] and the calculated value of Bi$_5$Ti$_5$O$_{17}$ using the finite-slab method [35].

B. Estimating $T_c$ of 2D-type superconductors.

We used the McMillan formula [36,37] to calculate the superconducting $T_c$.

$$T_c = \frac{\omega_{\log}}{1.2}\exp[\frac{-1.04(1+\lambda)}{\lambda - \mu^*(1+0.62\lambda)}] \quad (6)$$

with

$$\log(\omega_{\log}) = \frac{1}{\lambda}\sum_{\mathbf{q}\nu}\lambda_{\mathbf{q}\nu}\log\omega_{\mathbf{q}\nu} \quad (7)$$

and

$$\lambda = \sum_{\mathbf{q}\nu}\lambda_{\mathbf{q}\nu} = \sum_{\mathbf{q}\nu}\frac{\gamma_{\mathbf{q}\nu}}{2\pi D(\epsilon_F)\omega_{\mathbf{q}\nu}^2}, \quad (8)$$

where $\omega_{\mathbf{q}\nu}$ is the phonon frequency with the phonon momentum $\mathbf{q}$ and the branch $\nu$, $D(\epsilon_F)$ is the density of states per spin at the Fermi energy $\epsilon_F$, and $\gamma_{\mathbf{q}\nu}$ is the phonon linewidth,

$$\gamma_{\mathbf{q}\nu} = \frac{4\pi\omega_{\mathbf{q}\nu}}{N_{\mathbf{k}}}\sum_{nm\mathbf{k}}^{N_{\mathbf{k}}}|g_{nm}^{\nu}(\mathbf{k},\mathbf{k}+\mathbf{q})|^2\delta(\epsilon_{\mathbf{k}+\mathbf{q}m})\delta(\epsilon_{\mathbf{k}n}). \quad (9)$$

Here, $N_{\mathbf{k}}$ is the number of the $\mathbf{k}$-points, $\epsilon_{\mathbf{k}n}$ is the Kohn-Sham eigenvalue from $\epsilon_F$ with the electron momentum $\mathbf{k}$ and the band index $n$, and $g_{nm}^{\nu}(\mathbf{k},\mathbf{k}+\mathbf{q})$ is the electron-phonon coupling matrix:

$$g_{nm}^{\nu}(\mathbf{k},\mathbf{k}+\mathbf{q}) = \sum_{s}\frac{\mathbf{e}_{\mathbf{q}\nu}^{s}}{\sqrt{2M_s\omega_{\mathbf{q}\nu}}}\cdot\mathbf{d}_{mn}^{s}(\mathbf{k}+\mathbf{q},\mathbf{k}), \quad (10)$$

where $M_s$ is the mass of the $s$-th ion, $\mathbf{e}_{\mathbf{q}\nu}^{s}$ is the phonon eigenvector corresponds the eigenvalue $\omega_{\mathbf{q}\nu}$, and $\mathbf{d}_{mn}^{s}(\mathbf{k}+\mathbf{q},\mathbf{k})$ is the deformation potential matrix element,

$$\mathbf{d}_{mn}^{s}(\mathbf{k}+\mathbf{q},\mathbf{k}) = \left\langle u_{m\mathbf{k}+\mathbf{q}}\left|\frac{\delta v_{\mathrm{SCF}}}{\delta u_{\mathbf{q}s}}\right|u_{n\mathbf{k}}\right\rangle, \quad (11)$$

where $v_{\mathrm{SCF}}(\mathbf{r})$ is the periodic part of the self-consistent potential $V_{\mathrm{SCF}}(\mathbf{r})$, and $u_{\mathbf{q}s}$ is the displacement of the $s$-th ion with wave number $\mathbf{q}$.

The cylindrical Fermi-surface of 2D-type superconductors is disadvantageous for calculation of $T_c$, primarily due to poor convergence of the integration about the phonon-momentum $\mathbf{q}$ in Eqs. (7) and (8). In order to improve the convergence, we applied the Wannier-interpolation scheme for accurate interpolation of $\lambda_{\mathbf{q}\nu}$ and $\omega_{\mathbf{q}\nu}$ together with the optimized-tetrahedron method [38], which has no smearing parameters and leads to faster convergence than the linear-tetrahedron method. We elaborate this scheme in Sec II.C for $\lambda_{\mathbf{q}\nu}$ and Sec II.D for $\omega_{\mathbf{q}\nu}$.

C. Wannier-interpolation scheme for $\lambda_{\mathbf{q}\nu}$

The usual self-consistent density functional perturbation theory (DFPT) calculation [39] is formidable with ultra-dense grids due to the expensive calculation cost. Meanwhile, the Wannier-interpolation scheme for $\lambda_{\mathbf{q}\nu}$ enables the integration in Eqs. (7)-(9) over a dense Brillouin zone (BZ) sampling [40]. This scheme interpolates both electron eigenvalues $\epsilon_{\mathbf{k}n}$ and deformation potential matrix elements $d^s_{mn}(\mathbf{k}+\mathbf{q},\mathbf{k})$ from a coarse grid to a dense grid, keeping the same accuracy of the self-consistent DFPT calculation. In the original Wannier-interpolation scheme for $\lambda_{\mathbf{q}\nu}$, smearing methods are adopted to treat the integration with delta functions in Eq. (9), and requires convergence test about both BZ sampling and smearing parameters. Alternatively, we adopted the optimized-tetrahedron scheme [38] for the integration of Eq. (9) about $\mathbf{k}$ to remove the smearing parameters and to simplify the convergence test. However, as a trade-off, the divergence of $\lambda_{\mathbf{q}\nu}$ is not regularized while, in smearing scheme, the divergence is smeared out. The divergence at $\mathbf{q}=0$ can be avoided by using half a grid interval shifted BZ mesh, but the divergence at $\mathbf{q}=2\mathbf{k}_\text{F}$ cannot be avoided. To regularize the divergence, we resort to the original formula of Eq. (9), the non-adiabetic counterpart of Eq. (9):[41–43]

$$\tilde{\gamma}_{\mathbf{q}\nu} = \frac{4\pi}{N_{\mathbf{k}}} \sum_{nm\mathbf{k}}^{N_{\mathbf{k}}} |g_{nm}^{\nu}(\mathbf{k},\mathbf{k}+\mathbf{q})|^2 (f_{\mathbf{k}n} - f_{\mathbf{k}+\mathbf{q}m}) \delta(\epsilon_{\mathbf{k}+\mathbf{q}m} - \epsilon_{\mathbf{k}n} - \omega_{\mathbf{q}\nu}), \tag{12}$$

where $f_{\mathbf{k}n}$ is the occupation number of the state with the eigenvalue $\epsilon_{\mathbf{k}n}$. This form is so-called the full width half maximum (FWHM) linewidth, and have an upper bound $\tilde{\gamma}_{\mathbf{q}\nu} < 4\pi \max|g|^2 D(\epsilon_F)$. Under the adiabatic limit, i.e. $\omega_{\mathbf{q}\nu}/\epsilon_F \ll 1$, $\tilde{\lambda} = \sum_{\mathbf{q}\nu} \frac{\tilde{\gamma}_{\mathbf{q}\nu}}{2\pi D(\epsilon_F)\omega_{\mathbf{q}\nu}^2} \to \lambda$, regardless of $\gamma_{\mathbf{q}\nu}$ diverging or not (See Appendix A).

D. Wannier-interpolation scheme for $\omega_{\mathbf{q}\nu}$ or dynamical matrix

Typical 3D metals have smooth phonon dispersion that corresponds to a short-range interatomic-force-constant matrix in real-space. Such smooth dispersion ensures the accuracy of Fourier-interpolation of the dynamical matrix in reciprocal-space, $C_{sr}(\mathbf{q})$ of the *s*-th and *r*-th atom. However, in low-dimensional metals, the Kohn anomaly appears, making the Fourier-interpolation break down. We adopted a Wannier-interpolation scheme for $\omega_{\mathbf{q}\nu}$ developed by M. Calandra [44] to reproduce the Kohn-anomaly. Here, we simplified the Calandra-interpolation scheme to improve the convergence by combining the optimized-tetrahedron method. The Calandra-interpolation is a transformation of a dynamical matrix from a coarse grid $\mathbf{q}^c$ to a dense grid $\mathbf{q}^d$ and from a moderate smearing parameter $T_{\text{ph}}$ to a tiny smearing parameter $T_0$, i.e. $C_{sr}(\mathbf{q}^c, T_{\text{ph}}) \to C_{sr}(\mathbf{q}^d, T_0)$. The smearing parameter should be reduced to reproduce the Kohn-anomaly because a large smearing parameter smoothes out the Kohn-anomaly as the temperature effect on the Kohn-anomaly. The key idea of the Calandra-interpolation is to separate the dynamical matrix into the analytical part and the non-analytical part, similar to the correction for longitudinal-optical (LO) and transverse-optical (TO) modes splitting in polar materials [45]. The analytical part is corresponding to the bare phonon, and the non-analytical part indicates the screening effect induced by the electron-phonon coupling,

$$C_{sr}(\mathbf{q}^d, T_0) = C_{sr}^{\text{bare}}(\mathbf{q}^d) + C_{sr}^{\text{el-ph}}(\mathbf{q}^d, T_0). \tag{13}$$

The analytical part $C_{sr}^{\text{bare}}(\mathbf{q}^d)$ on the dense grid is obtained by Fourier-interpolation scheme and the non-analytical part $C_{sr}^{\text{el-ph}}(\mathbf{q}^d, T_0)$ is obtained by the Wannier-interpolation scheme. The analytical part of the dynamical matrix on the coarse grid is given as

$$C_{sr}^{\text{bare}}(\mathbf{q}^c) \equiv C_{sr}\left(\mathbf{q}^c, T_\infty\right) = C_{sr}(\mathbf{q}^c, T_{\text{ph}}) - \Lambda_{sr}(\mathbf{q}^c, T_{\text{ph}}) + \Lambda_{sr}\left(\mathbf{q}^c, T_\infty\right), \tag{14}$$

where the $T_\infty$ is a smearing parameter large enough to make $C_{sr}^{\text{bare}}(\mathbf{q}^c)$ smooth to ensure accurate Fourier-interpolation from $C_{sr}^{\text{bare}}(\mathbf{q}^c)$ to $C_{sr}^{\text{bare}}(\mathbf{q}^d)$, and

$$\Lambda_{sr}(\mathbf{q}, T_\alpha) = \frac{2}{N_\mathbf{k}(T_\alpha)} \sum_{nm\mathbf{k}}^{N_\mathbf{k}(T_\alpha)} \frac{f_{\mathbf{k}n}(T_\alpha) - f_{\mathbf{k+q}m}(T_\alpha)}{\epsilon_{\mathbf{k}n} - \epsilon_{\mathbf{k+q}m} + i\eta} d^s_{nm}(\mathbf{k}, \mathbf{k+q}) d^r_{mn}(\mathbf{k+q}, \mathbf{k}), \tag{15}$$

where $N_\mathbf{k}(T_\alpha)$ is the number of $\mathbf{k}$-points which ensures the convergence of the summation for a smearing parameter $T_\alpha$, and $f_{\mathbf{k}n}(T_\alpha)$ is the smeared occupation number of states with $\epsilon_{\mathbf{k}n}$ eigen-energy. $\eta$ is an infinitesimal value. Next, $C_{sr}^{\text{bare}}(\mathbf{q}^d)$ in the dense grid is obtained by Fourier-interpolation of $C_{sr}^{\text{bare}}(\mathbf{q}^c)$ in the coarse grid. Finally, the non-analytic part of the dynamical matrix on the dense grid is given as

$$C_{sr}^{\text{el-ph}}(\mathbf{q}^d, T_0) \equiv \Lambda_{sr}(\mathbf{q}^d, T_0) - \Lambda_{sr}\left(\mathbf{q}^d, T_\infty\right). \tag{16}$$

In order to calculate Eq. (16) in the dense grid, the electron eigenvalues $\epsilon_{\mathbf{k}n}$ and deformation potential matrix element $d^s_{mn}(\mathbf{k+q}, \mathbf{k})$ are Wannier-interpolated from the coarse grid.

Here, we simplify this original scheme by taking the limit $T_\infty \to \infty$ and $T_0 \to 0$. Under this limit, $\Lambda_{sr}\left(\mathbf{q}, T_\infty\right)$ vanishes because $f_{\mathbf{k}i}\left(T_\infty\right) - f_{\mathbf{k+q}i}\left(T_\infty\right)$ goes to zero, and $\Lambda_{sr}(\mathbf{q}^d, T_0)$ becomes the tetrahedron-integration version $\Lambda_{sr}^{\text{tet}}(\mathbf{q}^d)$. The above formulas Eqs. (13), (14), and

(16) are now replaced by

$$C_{sr}^{\text{tet}}(\mathbf{q}^d) = C_{sr}^{\text{bare}}(\mathbf{q}^d) + C_{sr}^{\text{el-ph}}(\mathbf{q}^d), \qquad (17)$$

$$C_{sr}^{\text{bare}}(\mathbf{q}^c) = C_{sr}(\mathbf{q}^c, T_{\text{ph}}) - \Lambda_{sr}(\mathbf{q}^c, T_{\text{ph}}), \qquad (18)$$

$$C_{sr}^{\text{el-ph}}(\mathbf{q}^d) = \Lambda_{sr}^{\text{tet}}(\mathbf{q}^d) \qquad (19)$$

with a tetrahedron version of Eq. (15):

$$\Lambda_{sr}^{\text{tet}}(\mathbf{q}) = \frac{2}{N_\mathbf{k}} \sum_{nm\mathbf{k}}^{N_\mathbf{k}} \frac{f_{\mathbf{k}n} - f_{\mathbf{k+q}m}}{\epsilon_{\mathbf{k}n} - \epsilon_{\mathbf{k+q}m} + i\eta} d_{nm}^s(\mathbf{k}, \mathbf{k+q}) d_{mn}^r(\mathbf{k+q}, \mathbf{k}). \qquad (20)$$

**III. COMPUTATIONAL DETAILS**

Electronic and dynamical properties are obtained using the QUANTUM-ESPRESSO [46] package with the optimized norm-conserving Vanderbilt (ONCV) pseudopotentials [47] and the revised Perdew–Burke–Ernzerhof exchange-correlation functional (PBEsol) [48]. The cut-off energy is 100 Ry for the planewave expansion of wavefunctions. The structures are optimized on an 8×8×6 **k**-grid using the optimized-tetrahedron method [38].

The spontaneous polarization is obtained using the modified Berry phase scheme. We generated an artificial adiabatic path to preserve the 2D-like electronic structure near the Fermi-level because the 2D property vanishes in optimized structures near the centrosymmetric phase. For the artificial structure of the centrosymmetric phase, we increased the **a** and **b** lattice parameters and decreased **c**. DFT+U scheme is also used as in Refs [27,49] to prevent the $O_{ap}$-$2p$ states from approaching the Fermi-level by increasing the onsite energy of $O_{ap}$-$2p$. We interpolated both lattice parameters and atomic positions from the structure with negative polarization to the artificial centrosymmetric structure and to the structure with positive polarization. The U value is turned on where the $O_{ap}$-$2p$ states tend to cross the Fermi-level. This procedure is justified because the polarization only depends on the first and the last structure. Moreover, the supercells are used to keep the number of electrons integer in the unit cell because the Berry phase scheme is developed on the integer number of electrons, i.e., 1×1×4 supercell for $n_H$=0.25 (holes per primitive cell) doping, 1×1×5 supercell for $n_H$=0.20 doping, and 1×1×6 supercell for $n_H$=0.167 doping.

The electron-phonon coupling matrix is obtained using DFPT on a coarse $N_\mathbf{k}(T_\mathrm{ph})$=12×12×6 **k**-grid and 6×6×3 **q**-grid using the Gaussian smearing method with smearing width $T_\mathrm{ph}$=0.02 Ry. The electronic eigenvalues, the deformation potential matrix, and the electron-phonon coupling matrix are interpolated on a dense $N_\mathbf{k}$=192×192×6 **k** and **q**-grid using the Wannier-interpolation scheme implemented in the electron-phonon coupling using Wannier functions (EPW) package [50]. Phonon frequencies and $T_c$ are calculated on the same dense grid using the scheme described at Secs. II.B, II.C, and II.D with Coulomb-pseudopotential $\mu^*$=0.16. To simulate the hole doping, electrons are removed from the system with compensating background charges added to maintain the charge neutrality.

**IV. RESULTS AND DISCUSSION**

Fig. 2(a) shows the polar displacement of $O_{ap}$ atom in the non-centrosymmetric (non-CS) phase as the hole-doping concentration is increased. The polar displacement is robust up to $n_H$=0.3 doping in $KTiO_2H$, and even over $n_H$>0.4 doping in $RbTiO_2H$ and $CsTiO_2H$. This result demonstrates two features of the polar distortion in the oxyhydrides; first, long-range electrostatic interaction, which is sensitive to the doping, gives little or no contribution to the polar distortion, whereas local bonding of $O_{ap}$ and Ti atoms gives the major contribution. Second, the polar distortion is weakly coupled to the itinerant hole carriers, as we expect from the electronic band structure in Fig. 1(b). This weak coupling condition persists until the narrow bands in the Γ-Y-T-Z plane from $O_{ap}$-2$p$ orbitals are unoccupied. In $KTiO_2H$, at $n_H$=0.3 doping, the apical oxygen $O_{ap}$-2$p$ states start to be unoccupied, and the polar distortion suddenly vanishes. In RbTiO$_2$H and CsTiO$_2$H, meanwhile, the narrow bands in the X-S-R-U plane from in-plane oxygen O$_{ip}$-2$p$ orbitals start to be unoccupied slightly earlier than the O$_{ap}$-2$p$ states [28], and the polar distortion persists at even higher doping concentrations than in KTiO$_2$H.

Besides, the zero-line in Fig. 2(a) indicates that the CS phase also persists as a local minimum structure. Although both CS and non-CS phases are dynamically stable, the latter becomes energetically less stable than the former with increasing the doping concentration, and the phase transition to the CS structure occurs at a critical doping concentration of $n_H$=0.08, 0.2, and 0.37 for KTiO$_2$H, RbTiO$_2$H, and CsTiO$_2$H, respectively. Therefore, in the oxyhydrides polar distortion persists under hole doping up to the critical concentrations.

Using the modified Berry phase scheme, we next calculated the spontaneous electric polarizations under hole-doping conditions [Fig. 2(b)]. Interestingly, the polarization is also resistant to the hole doping as the polar distortion. This robust polarization is attributed to incomplete screening by itinerant-holes $P_e^{hp}$ ($P^{total} - P^{bare}$) of the bare polarization $P^{bare}$. In KTiO$_2$H with $n_H$=0.2, $P^{bare}$ is about ~4.5 $e$Å per primitive cell (or 100 μC/cm$^2$), which is comparable to that of PbTiO$_3$, 50~100 μC/cm$^2$, and $P_e^{hp}$ is only about ~0.11 $e$Å due to the confinement of the hole carriers along the **c**-axis. Fig. 3(a) and (b) show that the hole carriers are localized mostly at H and O$_{ip}$ sites, and barely on O$_{ap}$ atoms because O$_{ap}$-2$p$ states are almost occupied at least up to $n_H$=0.3 as discussed above. Fig. 3(c) shows the response of the hole carriers upon the shift of O$_{ap}$ atom by $\Delta = 0.5$ Å. The redistribution of the hole carriers on the H-Ti-O$_{ip}$ plane is even smaller than the redistribution of dilute hole carriers at O$_{ap}$ sites, corresponding to the weak screening effect. Despite the non-zero spontaneous polarization, the ferroelectric switching may not be feasible due to the large displacement of about 1 Å, which is about ten times larger than that in PbTiO$_3$.

Fig. 4(a) shows the phonon dispersion of the non-CS phase and its modulation as the hole-doping concentration is increased. The non-CS phase of KTiO$_2$H is dynamically stable up to $n_H$~0.20, where the imaginary frequency modes start to emerge. The atomic projection of phonon modes shows the contrasting behavior for O$_{ap}$ and H vibrational modes. Due to the weak coupling between the polar displacement and the itinerant-hole [Fig. 2(a)], O$_{ap}$ vibrational modes are insensitive to the hole-doping without softening or hardening. On the other hand, H vibrational modes are sensitive to the hole-doping; high-frequency stretching modes [Fig. 2(c)] become hardened and less dispersive along Γ-Y line, and low-frequency bending modes [Fig. 2(d)] become softened and more dispersive along Γ-X line.

Such sensitive response of phonon dispersion to the hole doping usually reflects strong electron-phonon coupling, implying a possibility of phonon-mediated superconductivity. Indeed, the electron-phonon coupling linewidth is sizable at both H-stretching and H-bending modes [Fig. 4(b)]. The hardening and less dispersion of H-stretching mode, which may seem apparently inconsistent with strong electron-phonon coupling, is dominantly caused by the electrostatic potentials. We checked the contribution of both long-range and short-range electrostatic potentials to the phonon frequencies by calculating the LO-TO phonon splitting and the onsite dipole-dipole interaction term of the interatomic-force-constant matrix in real-space. For simplicity, we estimated the LO-TO splitting [45] and the onsite term of H-stretching mode relative to those of H-bending mode. The ratio of LO-TO splitting of H-stretching mode $\Delta\omega^2_{H-s}$ to H-bending mode $\Delta\omega^2_{H-b}$ is

$$R^{LO-TO} = \frac{\Delta\omega^2_{H-s}}{\Delta\omega^2_{H-b}} = \left(\frac{Z^*_{Hyy}}{Z^*_{Hxx}}\right)^2, \qquad (21)$$

where $Z^*_H$ is the born effective charge tensor of hydrogen anion. For the pristine KTiO$_2$H, the ratio is about 13.4 using the values of $Z^*_H$ elements in Ref. [28]. The large ratio indicates a large contribution of long-range electrostatic potential to the H-stretching mode. It explains the sensitive reduction of the dispersion width at higher hole-doping concentration due to the screening of long-range electrostatic potentials.

Similarly, the ratio of the onsite dipole-dipole interaction term of H-stretching mode $C_{\text{H}y\text{H}y}^{\text{DD}}(0,0)$ to H-bending mode $C_{\text{H}x\text{H}x}^{\text{DD}}(0,0)$ is obtained using the Gonze's formalism [45],

$$R^{00} = \frac{C_{\text{H}y\text{H}y}^{\text{DD}}(0,0)}{C_{\text{H}x\text{H}x}^{\text{DD}}(0,0)} = -2\frac{Z_{\text{H}yy}^{*}Z_{\text{Ti}yy}^{*}/\varepsilon_{yy}^{\infty}}{Z_{\text{H}xx}^{*}Z_{\text{Ti}xx}^{*}/\varepsilon_{xx}^{\infty}}, \quad (22)$$

where $\varepsilon^{\infty}$ is the electric contribution of the dielectric permittivity tensor. The ratio is about $-4.1$ in pristine KTiO$_2$H. The negative sign reflects concave curvature of the short-range electrostatic potential of H anion about y-direction displacement of H atom. Therefore, the screening of short-range interaction induces the overall hardening of the H-stretching branch. This effect predominates the softening by the electron-phonon coupling.

The phonon softening of H-bending modes with a kink at $\mathbf{q} = 2\mathbf{k}_F$ is a signature of the Kohn anomaly due to the electron-phonon coupling in low-dimensional metals. The acute and sharp Kohn anomaly at higher doping concentrations is a characteristic feature of 1D metals. Interestingly, there is no nesting at all in the Fermi surface of KTiO$_2$H with $n_H$=0.2 [Fig. 4(e)]. Usual 2D-type Kohn anomaly has muffin-tin shape as in MgB$_2$ [51], and becomes broader rather than deeper at high hole-doping concentration [52] as observed in hole-doped graphane (C$_2$H$_2$) [30]. 1D-type Kohn anomaly also occurs in graphene on Pt(111) [53], which was attributed to the linear dispersion of the energy bands. In terms of the non-parabolic dispersion effect, we compared the electronic band structure of KTiO$_2$H with its quadratic fitting at Y point [Fig. 5(a)]. The deviation from the quadratic fitting $\alpha k^2$ reflects a degree of quartic term, $\beta k^4$. In Fig. 5(b) and 5(c), we plotted the Lindhard functions $\chi(\mathbf{q})$ using the DFT band structure and its quadratic fitting as the Fermi-level is shifted. The $\chi(\mathbf{q})$ from the band structure exhibits the same feature of the Kohn anomaly along the $\Gamma$-X direction in Fig. 4(a), whereas the $\chi(\mathbf{q})$ from the quadratic fitting shows the typical 2D-type Kohn anomaly. This comparison suggests that the quartic term may induce a sharp Kohn anomaly in 2D-type metals. We elaborated this statement analytically in Appendix B. The $\chi(\mathbf{q})$ from the band structure shows a sharp peak also along the $\Gamma$-Y direction, and corresponding softening is expected on the H-stretching branch, a LO mode, but the hardening by the electrostatic effect predominates the softening by the electron-phonon coupling. Therefore the Kohn anomaly along the $\Gamma$-Y direction shows a rather blunt kink on the H-stretching branch.

Finally, we investigated the superconductivity in the hydroxides. As the hydrogen vibrational modes give a major contribution to the electron-phonon coupling, the light mass of hydrogen can advantageous for superconductivity through both $\omega_{\log} \propto \frac{1}{\sqrt{m}}$ and $\lambda \sim \frac{D(\epsilon_F)\langle d\rangle^2}{m\langle\omega\rangle^2} \propto \frac{1}{m\langle\omega\rangle^2}$. We found that the superconducting transition temperature $T_c$ is well below 1 K at low doping concentration ($n_H \leq 0.16$) despite the sizable $D(\epsilon_F)$ of 2D electronic bands and significant hydrogen contribution. Such loss of superconductivity reflects the dominant ionic-bonding character in $ATiO_2H$ and thus weak deformation potential. In the intermediate doping range ($0.16 \leq n_H < \sim 0.20$), the softening in the phonon modes by the quartic term in electrnoic bands reduces the $\omega_{\log}$ but highly enhances $\lambda$, leading to sizable $T_c$. Above $n_H \sim 0.20$, phonon frequencies turn into imaginary and the charge density wave corresponding to the hydrogen-dimerization-like distortion develops. The behavior of $T_c$ near the structural instability accompanied by sharp softening is also observed in bulk S, Se, and, Te under pressure [54,55].

Fig. 6(b) summarizes our main results. $ATiO_2H$ (A=K, Rb, Cs) can have spontaneous polarization under heavily hole-doped conditions. The superconductivity can coexist with the polarization near the hole doping level of $n_H \sim 0.2$ in $RbTiO_2H$ and $CsTiO_2H$. Above $n_H \sim 0.2$, the CDW phase coexists with the polarization. Substitution of Ti with Sc or In can be a practical way for hole doping as in SrTiO$_3$ [56,57]. We note that exotic superconductivity such as unusual paring or magneto-electric effect may be minor as the spin-orbit coupling (SOC) is insignificant in ATiO$_2$H. The band splitting by the SOC is only about 2 meV. The coexistence of spontaneous polarization and superconductivity in ATiO$_2$H allows the coupling between external electric field and superconductivity, another type of magneto-electric effect similar to that in ferroelectric/superconductor heterostructures [58].

## V. CONCLUSION

We investigated the crystal structure, spontaneous polarization, and superconductivity in perovskite-type hydroxides ATiO$_2$H (A=K, Rb, Cs). We used a modified Berry phase scheme for estimating the polarization for 2D-type metals, and improved Wannier-interpolation scheme for the electron-phonon coupling constant and phonon frequencies combining the optimized-tetrahedron method. We showed that the oxyhydrides can have persistent polarization up to sizable hole-doping concentrations. We found that the superconductivity can coexist with the polarization at optimal doping concentration in RbTiO$_2$H and CsTiO$_2$H. Therefore, the magneto-electric effect, crucial in spintronics or fluxtronics, could present in these compounds. Our findings indicate one promising route to design of polar metals with phonon-mediated superconductivity.

## ACKNOWLEDGMENTS

This study was supported by the National Research Foundation of Korea (NRF) grant (No.2018R1A5A6075964) funded by the Korea government (MSIT). Supercomputing resources including technical supports were provided by Supercomputing Center, Korea Institute of Science and Technology Information (Contract No. KSC-2017-C3-0070).

**Appendix A:** Proof of $\tilde{\lambda} \to \lambda$ under $\omega_{\mathbf{q}\nu}/\epsilon_F \ll 1$ limit

Rewriting the Eq. (12),

$$\tilde{\gamma}_{\mathbf{q}\nu} = \frac{4\pi}{N_{\mathbf{k}}} \sum_{nm\mathbf{k}}^{N_{\mathbf{k}}} |g_{nm}^{\nu}(\mathbf{k}, \mathbf{k}+\mathbf{q})|^2 (H(\epsilon_{\mathbf{k}n} + \omega_{\mathbf{q}\nu}) - H(\epsilon_{\mathbf{k}n})) \delta(\epsilon_{\mathbf{k}+\mathbf{q}m} - \epsilon_{\mathbf{k}n} - \omega_{\mathbf{q}\nu}), \quad \text{(A1)}$$

where the step function $H(\epsilon_{\mathbf{k}n}) = -f_{\mathbf{k}n}$.

Using $\int_a^b \delta(x) dx = H(b) - H(a)$,

$$\tilde{\gamma}_{\mathbf{q}\nu} = \int_0^{\omega_{\mathbf{q}\nu}} d\Omega \frac{4\pi}{N_{\mathbf{k}}} \sum_{nm\mathbf{k}}^{N_{\mathbf{k}}} |g_{nm}^{\nu}(\mathbf{k}, \mathbf{k}+\mathbf{q})|^2 \delta(\epsilon_{\mathbf{k}n} + \Omega) \delta(\epsilon_{\mathbf{k}+\mathbf{q}m} - \epsilon_{\mathbf{k}n} - \omega_{\mathbf{q}\nu}). \quad \text{(A2)}$$

Let $\Omega^{\max}$ ($\Omega^{\min}$) be a value of $\Omega$ ($\leq \omega_{\mathbf{q}\nu}$), which makes the integrand of Eq. (A2) maximum (minimum), and

$$\tilde{\gamma}_{\mathbf{q}\nu}^{>} \equiv \omega_{\mathbf{q}\nu} \times \frac{4\pi}{N_{\mathbf{k}}} \sum_{nm\mathbf{k}}^{N_{\mathbf{k}}} |g_{nm}^{\nu}(\mathbf{k}, \mathbf{k}+\mathbf{q})|^2 \delta(\epsilon_{\mathbf{k}n} + \Omega^{\max}) \delta(\epsilon_{\mathbf{k}+\mathbf{q}m} + \Omega^{\max} - \omega_{\mathbf{q}\nu}), \quad \text{(A3)}$$

$$\tilde{\gamma}_{\mathbf{q}\nu}^{<} \equiv \omega_{\mathbf{q}\nu} \times \frac{4\pi}{N_{\mathbf{k}}} \sum_{nm\mathbf{k}}^{N_{\mathbf{k}}} |g_{nm}^{\nu}(\mathbf{k}, \mathbf{k}+\mathbf{q})|^2 \delta(\epsilon_{\mathbf{k}n} + \Omega^{\min}) \delta(\epsilon_{\mathbf{k}+\mathbf{q}m} + \Omega^{\min} - \omega_{\mathbf{q}\nu}). \quad \text{(A4)}$$

Therefore,

$$\tilde{\gamma}_{\mathbf{q}\nu}^{<} \leq \tilde{\gamma}_{\mathbf{q}\nu} \leq \tilde{\gamma}_{\mathbf{q}\nu}^{>}. \quad \text{(A5)}$$

Under $\omega_{\mathbf{q}\nu}/\epsilon_F \ll 1$ limit, $\tilde{\gamma}_{\mathbf{q}\nu}^{>}$ and $\tilde{\gamma}_{\mathbf{q}\nu}^{<}$ approach $\gamma_{\mathbf{q}\nu}$, and $\tilde{\gamma}_{\mathbf{q}\nu} \to \gamma_{\mathbf{q}\nu}$ unless $\gamma_{\mathbf{q}\nu}$ diverges.

Meanwhile, the $\tilde{\lambda} = \sum_{q\nu} \frac{\tilde{\gamma}_{q\nu}}{2\pi D(\epsilon_F) \omega_{\mathbf{q}\nu}^2}$ satisfies

$$\tilde{\lambda}^{<} \leq \tilde{\lambda} \leq \tilde{\lambda}^{>} \quad \text{(A6)}$$

Under $\omega_{\mathbf{q}\nu}/\epsilon_F \ll 1$ limit, $\tilde{\lambda}^{>}$ and $\tilde{\lambda}^{<}$ approach $\lambda$, and thus $\tilde{\lambda} \to \lambda$, which is always valid due to the upper bound ($\lambda < \max|g|^2 D(\epsilon_F)^2$).

**Appendix B:** Analytical proof of the statement, *quartic term induces sharp Kohn anomaly in 2D-type metals*

A simple model can reproduce the sharp peak in the phonon dispersion. We assumed an isotropic quartic dispersion in 2D:

$$\epsilon_{\mathbf{k}} = -\frac{\hbar^2}{2m}(\mathbf{k}^2 - \alpha \mathbf{k}^4) - \epsilon_F, \tag{B1}$$

with Fermi level $\epsilon_F$ and Fermi momentum $k_F$. The Lindhard function $\chi(\mathbf{q})$ is expressed as

$$\chi(\mathbf{q}) = 2\int \frac{d^2\mathbf{k}}{(2\pi)^2} \frac{f_{\mathbf{k+q}} - f_{\mathbf{k}}}{\epsilon_{\mathbf{k+q}} - \epsilon_{\mathbf{k}}} = 2\int_{k<k_F} \frac{d^2\mathbf{k}}{(2\pi)^2}\left(\frac{1}{\epsilon_{\mathbf{k+q}} - \epsilon_{\mathbf{k}}} - \frac{1}{\epsilon_{\mathbf{k}} - \epsilon_{\mathbf{k-q}}}\right). \tag{B2}$$

If $\alpha k_F^2 \ll 1$,

$$-\chi(q \le 2k_F) = D(\epsilon_F) + \frac{\alpha m}{3\pi\hbar^2}q^2$$

$$-\chi(q > 2k_F) = D(\epsilon_F) + \frac{\alpha m}{3\pi\hbar^2}q^2 - \frac{m}{\pi\hbar^2}\left[\left(1 - \frac{4k_F^2}{q^2}\right)^{\frac{1}{2}} + \frac{\alpha}{3}q^2\left(1 - \frac{4k_F^2}{q^2}\right)^{\frac{3}{2}}\right], \tag{B3}$$

where $D(\epsilon_F)$ is the density of states at $\epsilon_F$. $D(\epsilon_F)$ is not a constant, but an increasing function about $k_F$.

$$D(\epsilon_F) = \frac{m}{\pi\hbar^2}(1 + 2\alpha k_F^2). \tag{B4}$$

The phonon softening is proportional to $\chi(q)$ (i.e. $\Delta\omega^2 \sim |g|^2 \chi(q)$). Therefore, the increasing $D(\epsilon_F)$ induces overall phonon softening, and the term $\frac{\alpha m}{3\pi\hbar^2}q^2$ in Eq. (B3) induces sharp softening proportional to the quartic term.

**FIGURES**

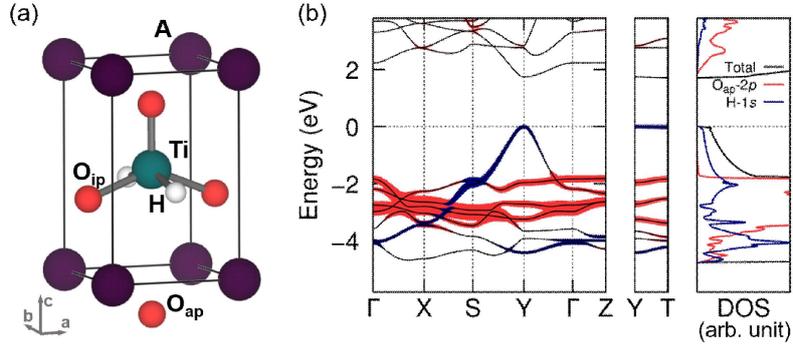

FIG. 1. (a) Crystal structure of ATiO$_2$H (A=K, Rb, Cs). O$_{ip}$ and O$_{ap}$ are in-plane and apical oxygen atoms, respectively. (b) Calculated electronic band structure of KTiO$_2$H with orbital projection to H-1$s$ orbital (in blue) and O$_{ap}$-2$p$ orbital (in red) as scaled to the linewidth. The total and partial density of states are also plotted in the right panel.

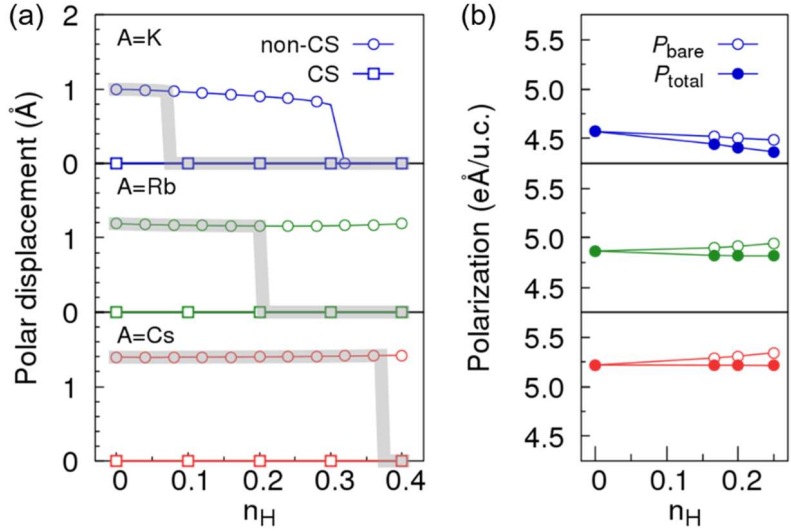

FIG. 2. (a) O$_{ap}$-A (A=K, Rb, Cs) relative displacement in non-CS phase ATiO$_2$H in terms of $n_H$ (holes per primitive cell). The zero displacement indicates that the CS phase is also a local-minimum structure. Thick lines in grey highlight the phase with a lower total energy. (b) The bare-polarization ($P^{\text{bare}} = P_{\text{ion}} + P_e^{\text{bare}}$) and the total-polarization ($P^{\text{total}} = P^{\text{bare}} - P_e^{\text{hp}}$) of the non-CS phase.

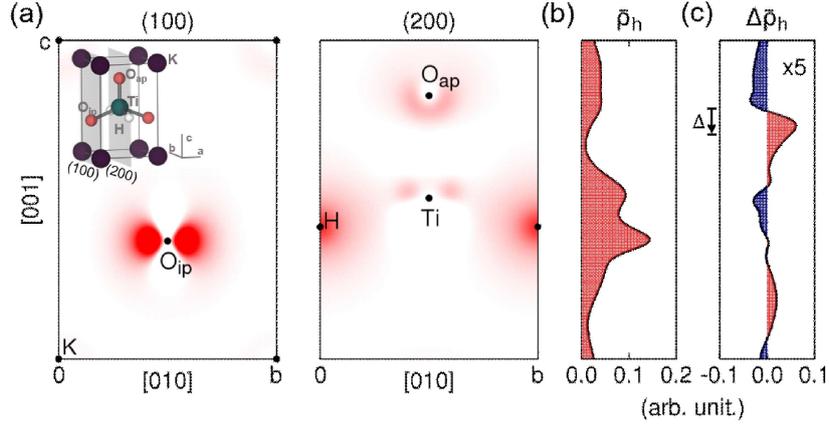

FIG. 3. (a) The hole-charge density plot in the (100)- and (200)-plane for $KTiO_2H$ with $n_H=0.2$ doping level. The inset shows the structure of $KTiO_2H$ with the lattice planes. (b) The planer-averaged hole-charge density and (c) its change due to the $\Delta=0.5$ Å downshift of $O_{ap}$-atom along the $c$-direction. The region in red (blue) indicates the excess of hole- (electron-) charge.

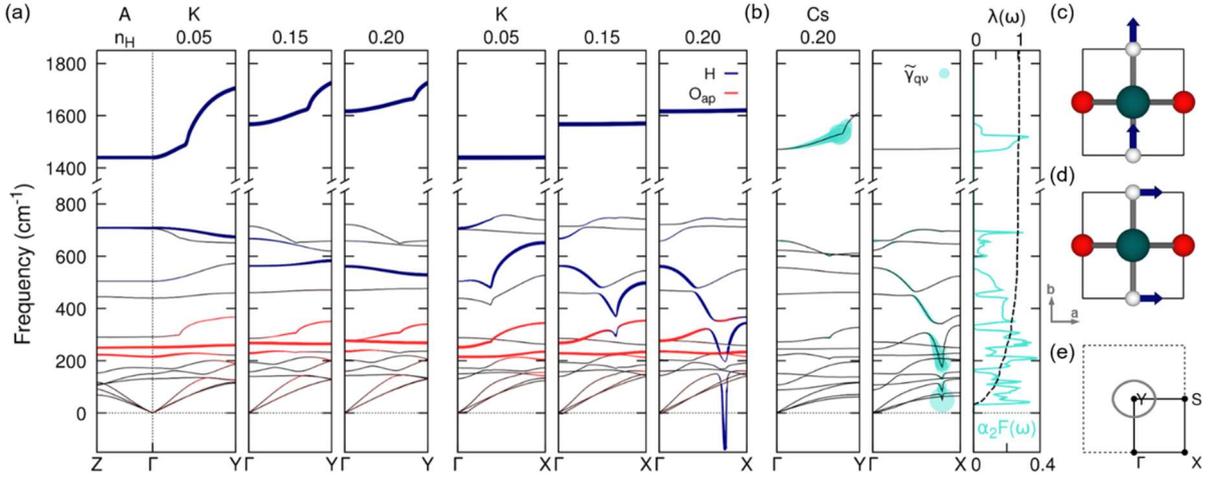

FIG. 4. (a) Calculated phonon dispersion of non-CS phase of hole-dope $KTiO_2H$ ($n_H$=0.05, 0.15, 0.20) with atomic projection to $O_{ap}$ (red line) and H (blue line) highlighted in proportion to the line width. (b) The electron-phonon coupling linewidth $\tilde{\gamma}_{qv}$ on the phonon dispersion and Eliashberg function $\alpha_2 F(\omega)$ with its integration $\lambda(\omega) = \int_0^\omega d\Omega\, \alpha_2 F(\Omega)/\Omega$ of $CsTiO_2H$ with $n_H$=0.20. Schematic illustration of hydrogen displacement for (c) stretching and (d) bending modes. (e) The Fermi-surface of $n_H$=0.20 doped $KTiO_2H$ on the $k_z$=0 plane.

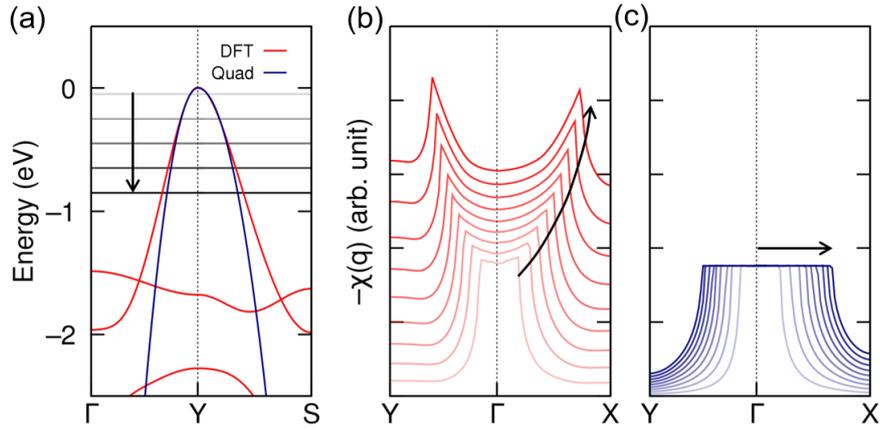

FIG. 5. (a) Calculated band structures of $KTiO_2H$ and their quadratic fitting curve (blue). The arrow highlights the shift of the Fermi level from $-0.05$ to $-0.85$ eV. (b) and (c), the Lindhard function $\chi(q)$ calculated from DFT band structure and its quadratic fitting, respectively, as the Fermi-level is shifted the same way in (a). The kink increases sharply in (b) where as it is only shifted toward X point in (c) as indicated by the arrows.

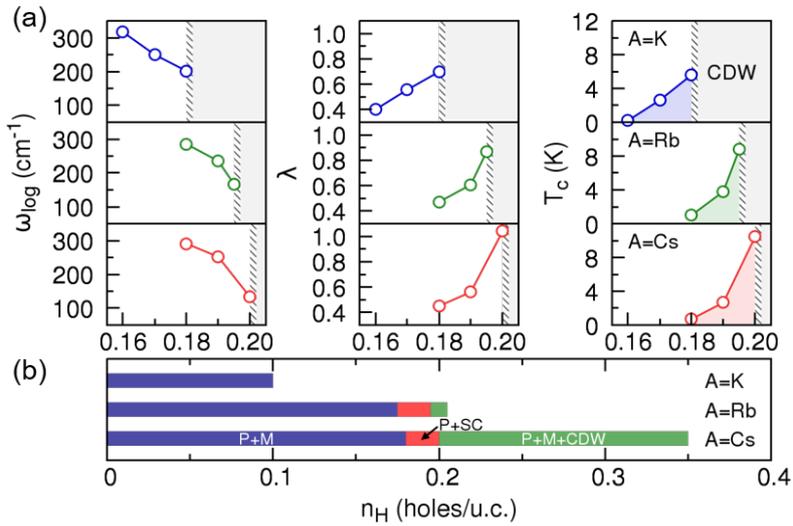

FIG. 6. (a) Averaged phonon frequency ($\omega_{\log}$), electron-phonon coupling constant ($\lambda$), and transition temperature ($T_c$) of $ATiO_2H$ (A=K, Rb, Cs) at different doping levels ($n_H$). (b) Coexistence of multiple phases in $ATiO_2H$ with doping levels ($n_H$) increased. P, non-zero spontaneous polarization; M, metallicity; SC, superconductivity; CDW, charge density wave.

28